\documentclass[sigconf]{acmart}
\AtBeginDocument{%
  }

\setcopyright{acmlicensed}
\copyrightyear{2018}
\acmYear{2018}
\acmDOI{XXXXXXX.XXXXXXX}

\acmJournal{JACM}
\acmVolume{37}
\acmNumber{4}
\acmArticle{111}
\acmMonth{8}

\usepackage[skins,breakable]{tcolorbox}
\begin{document}

\title{Ask before you Build: Rethinking AI-for-Good in Human Trafficking Interventions}

   
\author{Pratheeksha Nair}
\email{pratheeksha.nair@mail.mcgill.ca}
\affiliation{%
  \institution{McGill University, Mila-Quebec AI Institute}
  \city{Montreal}
  \country{Canada}
}

\author{Gabriel Lefebvre}
\email{gabriel.lefebvre@mail.mcgill.ca}
\affiliation{\institution{McGill University}
\city{Montreal}
\country{Canada}
}

\author{Sophia Garrel}
\email{sophiagarrel@gmail.com}
\affiliation{%
  \institution{Universite de Montreal, Mila-Quebec AI Institute}
  \city{Montreal}
  \country{Canada}
  }

\author{Maryam Molamohammadi}
\affiliation{%
 \institution{Mila-Quebec AI Institute}
 \city{Montreal}
 \country{Canada}
 }

\author{Reihaneh Rabbany}
\email{rrabba@cs.mcgill.ca}
\affiliation{%
  \institution{McGill University, Mila-Quebec AI Institute}
  \city{Montreal}
  \country{Canada}
  }

\renewcommand{\shortauthors}{Nair et al.}

\begin{abstract}
AI-for-good initiatives often rely on the assumption that technical interventions can resolve complex social problems. In the context of human trafficking (HT), such techno-solutionism risks oversimplifying exploitation, reinforcing power imbalances, and causing harm to the very communities AI claims to support. In this paper, we introduce the Radical Questioning (RQ) framework as a five-step, pre-project ethical assessment tool to critically evaluate whether AI should be built at all—especially in domains involving marginalized populations and entrenched systemic injustice. RQ does not replace principles-based ethics but precedes it, offering an upstream, deliberative space to confront assumptions, map power, and consider harms before design. Using a case study in AI for HT, we demonstrate how RQ reveals overlooked socio-cultural complexities and guides us away from surveillance-based interventions toward survivor-empowerment tools. While developed in the context of HT, RQ’s five-step structure can generalize to other domains—though the specific questions must be contextual. This paper situates RQ within a broader AI ethics philosophy that challenges instrumentalist norms and centers relational, reflexive responsibility.
\end{abstract}



\received{20 February 2007}
\received[revised]{12 March 2009}
\received[accepted]{5 June 2009}

\maketitle

\section{Introduction}

In recent years, the deployment of artificial intelligence (AI) in social good contexts has surged, driven by the belief that technology can offer effective solutions to complex societal challenges~\cite{green2019good}.
Most AI-for-good projects inherit a techno-solutionist mindset~\cite{heilinger2022ethics, riley2008engineering, schull2013folly, metcalf2019owning}—believing that complex social problems are reducible to datasets and solvable via automation. In domains like human trafficking (HT), this framing is not only inadequate but dangerous: it reifies harmful narratives, empowers surveillance actors, and silences those directly affected. 

Recent AI interventions targeting HT have largely focused on crime-sleuthing applications, including machine learning classifiers for escort ads~\cite{tnet,dubrawski2015leveraging,alvari2017semi,tong2017combating,li2023detecting,mensikova2018ensemble}, computer vision systems for identifying trafficking indicators in imagery~\cite{stylianou2019traffickcam} and geotags~\cite{bamigbade2024computer}, and network-mapping tools used by law enforcement and NGOs to detect trafficking “hotspots”~\cite{polaris}. These tools typically frame trafficking as a data problem—one to be solved through pattern recognition and predictive modeling. However, critics note that such approaches rest on narrow assumptions, prioritize enforcement over survivor well-being, and frequently ignore systemic root causes~\cite{musto2014trafficking,musto2020between,deeb2022ethical}. Studies have shown that these models can perpetuate racialized and gendered biases, misclassify consensual sex work as trafficking, and ultimately reinforce surveillance and carceral systems~\cite{brown2024policing}. We argue that these shortcomings calls for upstream, human-centered frameworks that interrogate the ethics of intervention before technical development begins.

Situating this upstream intervention within the existing ethical frameworks~\cite{leslie2019understanding,fetic2020principles,madaio2020co} — these center on principles like transparency and fairness — are reactive and compliance-oriented, and often implemented only after technical design has begun. These seldom ask a more fundamental question: Should we be building this at all?  

In this paper, we introduce Radical Questioning (RQ) as an upstream, pre-design ethics framework that foregrounds this very question. Developed through interdisciplinary collaboration in the HT context, RQ guides AI developers through critical reflection about the motivations, implications, and legitimacy of their proposed intervention. Unlike participatory design~\cite{bhalerao2022analyzing,heilinger2022ethics,deeb2022ethical} or risk assessments~\cite{human_rights_ai_impact_assessment}, RQ is agnostic to outcomes and rooted in philosophical deliberation~\cite{eubanks2018automating,benjamin2019race,fraser2009scales} — it does not prescribe what should be done, but how to interrogate why we believe something ought to be done in the first place.

RQ’s novelty lies in both its structural role—intervening before AI design—and its application to the hyper-contested, ethically fraught domain of HT, which provides a particularly acute example of how AI can exacerbate harm under the guise of social good. While developed in the HT context, RQ’s structure is generalizable to other settings where ethical complexity precedes technical feasibility.
Moreover, we call for a paradigm shift in AI ethics: to normalize pre-project ethical assessment as a core component of responsible AI practice, particularly in domains marked by contested definitions, systemic harms, and stakeholder asymmetries.

\section{Related Works}

Our framework builds on a growing body of research that critiques techno-solutionism and advocates for context-sensitive, justice-oriented approaches to AI design. Refusal-based frameworks such as Studying Up~\cite{studying_up} and the broader theory of data refusal~\cite{barabas2022refusal} challenge dominant problem framings by interrogating harmful assumptions and the positionality of power. Similarly, Design Justice~\cite{costanza2020design} and the heterogeneity framework~\cite{molamohammadi2023unraveling} emphasize centering marginalized voices and accounting for intersections of culture, law, and infrastructure. While RQ shares these commitments to power analysis and critical reflection, it extends them into a structured, iterative, and pre-design methodology that explicitly asks whether an AI system should be built at all.

Other scholars call for re-centering AI ethics on foundational moral questions~\cite{green2019good,heilinger2022ethics}, or critique how AI amplifies state surveillance and racialized carceral logics in the context of trafficking~\cite{musto2014trafficking,milivojevic2020freeing}. Complementary tools like RED (Rapid Ethical Deliberation)~\cite{steen2021method} provide procedural or risk-centered guidance~\cite{razi2021human} for high-stakes domains but typically assume a system is already in development.

Deliberation-focused approaches like the Situate AI Guidebook~\cite{situate_ai} and participatory frameworks~\cite{delgado2023participatory} support structured stakeholder engagement, while HCI contributions—such as trauma-informed computing~\cite{kelly2022trauma}, survivor-centered justice~\cite{rabaan2023survivor}, and digital-safety research protocols~\cite{bellini2024sok}—offer critical guidance for working with vulnerable populations. However, many of these operate downstream of the initial decision to build and rarely offer mechanisms to pause or reframe interventions entirely. RQ complements and deepens these efforts by introducing open-ended, upstream deliberation that challenges the necessity and legitimacy of technical intervention at its root.

In the HT domain, prior work has identified risks such as dataset bias, privacy violations, and the harms of over-policing, especially when AI tools are deployed without attention to socio-cultural complexity~\cite{deeb2022ethical,bhalerao2022analyzing}. The RQ framework addresses this gap by offering a pre-design process that surfaces these concerns early and critically, enabling more ethically sound decisions before systems are built. Rather than relying on fixed stakeholder roles or prescriptive checklists, RQ creates reflective space for foundational questioning—making it compatible with but not dependent on institutional or participatory structures. Within HT, it uniquely reshapes design trajectories by grounding intervention in relational ethics and moral deliberation.

\section{Radical Question Framework}

We introduce Radical Questioning (RQ) as a pre-design ethics framework developed through its application in the human trafficking (HT) domain. RQ refers to the practice of interrogating foundational assumptions that shape how problems are defined and solutions are pursued. Rather than asking how to build better AI tools for combating HT, RQ begins by asking why AI is the appropriate response at all—and who benefits or is harmed by its deployment. Questions such as “What does justice mean in this context?” and “Whose definitions are being used, and why?” invite critical reflection on the normative, political, and institutional stakes of AI-for-good projects.

The Radical Questioning (RQ) framework emerged through sustained engagement with survivor-led organizations and interdisciplinary collaborators, grounded in the legal and social complexities of the human trafficking (HT) domain. While context-specific in origin, RQ’s five-step process is broadly applicable to other ethically complex domains—so long as its questions are rooted in the specific realities of those fields. Designed for AI developers, researchers, policymakers, and interdisciplinary teams, RQ is especially valuable in high-stakes contexts where definitions are contested and harm is unevenly distributed. It guides not implementation, but reflection on whether an AI intervention should proceed at all.

Although not a participatory design method, RQ is grounded in relational accountability: its questions were co-shaped by individuals differently situated in relation to harm, power, and intervention. Rather than a checklist, RQ is a deliberative practice that invites open-ended inquiry, iterative reflection, and deep stakeholder engagement to resist premature ethical closure.

\begin{figure*}
    \centering
\includegraphics[width=0.6\linewidth]{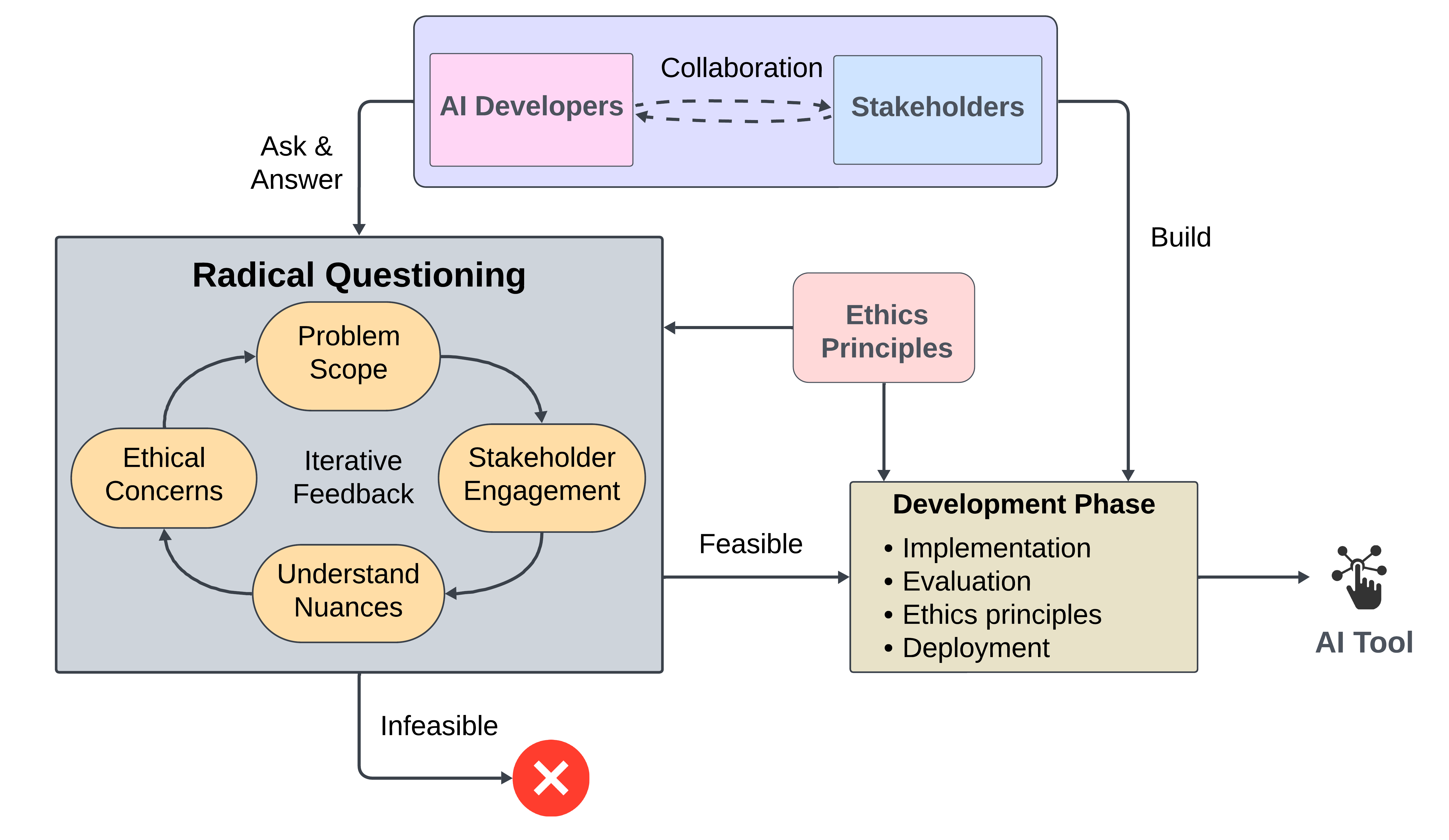}
    \caption{The proposed RQ framework is based on asking and answering radical questions through deliberative communication and collaboration between AI developers and stakeholders before development. If deemed ethically feasible, the tool is developed and if not, it is terminated.}
    \label{fig:proposed_framework}
\end{figure*}

Below, we present RQ’s five steps, accompanied by example questions developed and refined through our HT case study. Gray boxes in the following section highlight actual questions raised during the design process to illustrate how RQ unfolds in practice.We also present RQ diagrammatically in Figure \ref{fig:proposed_framework}.

\subsection{Step 1: Define the Scope of the Problem}

This step asks not what the technical problem is, but what the social issue being addressed actually means—and who gets to define it. In our HT case, the initial framing was: “How can we detect trafficking online using escort advertisements?” Yet this presumes that trafficking is legible to machine learning, that online ads are reliable indicators, and that detection is the right goal. By asking: Why are we framing HT in this way? Who benefits from this definition? we learned that much of the literature and tooling assumes a fixed, binary notion of exploitation.

HT laws in Canada, for example, are often invoked in ways that conflate sex work with exploitation, and are disproportionately used against migrant sex workers~\cite{konrad,brown2024policing}. Moreover, assumptions such as “pimps write the ads”~\cite{crotty2018red, bouche2015gendered} ignore survivor agency and may lead to harmful interventions. Such assumptions erase the agency of sex workers and legitimizes surveillance-based interventions that can put marginalized communities at further risk. This shifted our framing from “detection of trafficking” to “support for survivors.”

\begin{tcolorbox}[breakable,width=\linewidth, sharp corners=all, colback=white!95!black, top=-1pt, left=-0.5cm]
\begin{itemize}
    \item What is human trafficking and why is it problematic? 
    \item Why are we defining the problem in this particular way, and who benefits from this framing?
    \item Is this a problem that can be alleviated using AI? Is there a demand for solving this problem and who has raised the demand?
    \item Are there particular problems within the HT domain that we can focus on? What sort of resources are required for solving these problems? 
\end{itemize}
\end{tcolorbox}

\subsection{Step 2: Identify Stakeholders}

AI development often privileges institutional stakeholders—such as law enforcement or funders—while marginalizing the perspectives of those most affected. We asked: Who will be impacted by this tool? Who owns it? Are our designs centered around who holds power or the proximity to potential harm?

In the HT ecosystem, stakeholder perspectives diverge drastically. Law enforcement views tech as a tool for surveillance and arrest~\cite{tong2017combating,spotlight_cite}; NGOs often focus on victim identification~\cite{rodrigues2020legal,gab_pred}; sex worker organizations advocate for harm reduction and autonomy~\cite{brown2024policing}. Survivors themselves express complex, sometimes ambivalent relationships to justice and intervention~\cite{farrell2019failing, sterling2018we}. We found that involving survivors directly—not just via proxies—radically altered the design direction. This engagement also revealed logistical challenges: many survivor groups had justifiable distrust in institutional actors, and sustained trust-building was necessary.

\begin{tcolorbox}[breakable,width=\linewidth, sharp corners=all, colback=white!95!black, top=-1pt, left=-0.5cm]
\begin{itemize}
\item Who will be the end user of our tool? Who will own the tool and ensure its proper use? 
    \item How do the different stakeholders implicated in the project understand the function and limits of the criminal law/and technology?
    \item Are we privileging certain stakeholders (e.g., law enforcement) over others (e.g., survivors), and what are the implications?
    \item Have we involved those directly affected by the tool, including marginalized voices, in meaningful ways?
    \item Do we have sufficient resources and expertise to engage meaningfully with multiple stakeholders?
\end{itemize}
\end{tcolorbox}

\subsection{Step 3: Understand Contextual Nuance}

Even when stakeholders are engaged, ethical blind spots remain if practitioners don't grapple with domain complexity. In HT, notions of justice, consent, and exploitation are far from universal. Some survivors do not want to press charges or be rescued; others fear the criminal justice system more than their traffickers. Even identifying “risk factors” (e.g., ethnic markers such as ``black'', ``asian'', extreme services like ``BDSM'', "sodomy"', sexual orientation indicators like ``queer'' or ``trans'', or language cues like poor English indicating foreigners)~\cite{giommoni2021identifying} risks reinforcing racial and gendered surveillance.

A particularly salient example is the “chilling effect”~\cite{schauer1978fear, young2023hansman} on sex workers caused by AI-based ad monitoring: fear of criminalization leads to self-censorship and economic insecurity. RQ helped us avoid replicating such harm by reframing success not as catching traffickers, but as supporting survivors’ pursuit of justice on their terms.

\begin{tcolorbox}[breakable,width=\linewidth, sharp corners=all, colback=white!95!black, top=-1pt, left=-0.5cm]
\begin{itemize}
    \item Have we uncovered the underlying complexities of the problem? What do these complexities mean for each stakeholder?
    \item How do different stakeholders perceive the problem we aim to address and do they agree with our approach of addressing it?
    \item How do we measure the success of the tool, and who decides what success looks like?
    \item Is our metric of success causing unintended harms to any stakeholders?
\end{itemize}
\end{tcolorbox}

\subsection{Step 4: Map Ethical Concerns}

This step surfaces questions around accountability, privacy, fairness, legitimacy, and incentives. We asked: What does fairness mean in this domain? Who defines accountability? We found that ground truths for “exploitation” are based on legal and social interpretations that evolve—and that AI tools trained on them can reinforce biased narratives~\cite{musto2014trafficking,giommoni2021identifying}. Equally, developers often escape accountability, while the tool becomes a mechanism for state power. RQ helped surface the disconnect between legal structures and social legitimacy. Additionally, in HT, legal compliance alone is insufficient for legitimacy~\cite{konrad}. For instance, privacy laws may permit certain uses of ``public'' data, but affected communities may still experience real harm. For example, an AI tool that flags “suspicious” ads can generate false positives, prompting police surveillance of consensual sex workers and retraumatizing victims.

\begin{tcolorbox}[breakable,width=\linewidth, sharp corners=all, colback=white!95!black, top=-1pt, left=-0.5cm]
\begin{itemize}
    \item What measures are in place to assess the tool’s fairness, accuracy, and social impact? Who decides what is fair and accurate? Which stakeholders were involved in discussing these metrics?
    \item How to choose who needs to take accountability for the tool? What does accountability mean to those involved in the domain? 
    \item Is simple conformity to law and constitutional requirements a sufficient bases for legitimacy of the initiative? 
    \item What does privacy mean for those affected by the problem and the tool?
    \item What is the composition of the team behind the tool and what are their motives and incentives?
\end{itemize}
\end{tcolorbox}

\subsection{Step 5: Iterate with Feedback}

Finally, RQ requires continuous stakeholder input, especially from vulnerable groups. But feedback is not merely about checking boxes—it must be deliberative, responsive, and open to halting a project entirely. We asked: Are we truly integrating critique? Are we willing to stop if harm outweighs benefit?

Our team participated in trauma-informed workshops with engaged with survivor-led organizations, while remaining attentive to the risks of retraumatization~\cite{witkin2018trauma}. We built an advisory board including survivors to oversee the tool’s development and function. Their feedback prompted design pivots—including reducing automation, foregrounding consent, and focusing on evidence preservation (not detection). Without this feedback, the project might have reinforced the very systems it sought to challenge.

\begin{tcolorbox}[breakable,width=\linewidth, sharp corners=all, colback=white!95!black, top=-1pt, left=-0.5cm]
\begin{itemize}
    \item Are there systems in place to obtain continuous feedback from stakeholders/survivors? How to be mindful of the risks of retraumatization? 
    \item Are we actively considering critiques that contradict our initial goals, or are we selectively responding to feedback that supports our pre-existing views?
    \item Are we truly acting on the feedback, or simply acknowledging it to maintain the appearance of responsiveness?
    \item Are there stakeholders whose feedback we are dismissing because it challenges the feasibility or goals of the project?
\end{itemize}
\end{tcolorbox}

\section{Challenges and Limitations}

While RQ proved valuable in this project, we also encountered several note-worthy challenges. 1) RQ is not prescriptive. Its strength lies in surfacing ethical dilemmas, but it does not dictate clear answers or technical implementation paths. This can be challenging for practitioners seeking actionable outcomes in time-constrained or product-driven environments. 2) The effectiveness of RQ hinges on genuine stakeholder engagement. In practice, building trust with affected communities—especially those historically marginalized or criminalized—requires time, resources, and institutional support. In our case, it took sustained outreach to survivor-led organizations before meaningful consultation could begin. In other contexts, such access may be limited or mediated through gatekeepers. 3) The outcomes of RQ are contingent on the positionality and openness of the development team. If teams are unwilling to alter project goals or confront uncomfortable truths, RQ risks becoming performative. Moreover, RQ requires developers to take on ethical responsibility themselves rather than outsourcing it to compliance checklists or institutional norms. 4) While the RQ process is transferable, the specific questions and concerns are not. Applying RQ to new domains will require re-grounding in context-specific histories, power dynamics, and social imaginaries. The framework is only as effective as the depth and sincerity of the questions asked.

\section{Conclusions and Takeaways}

Applying the Radical Questioning (RQ) framework to human trafficking (HT) fundamentally reshaped our project. What began as an AI-for-good intervention—automated detection of trafficking in online ads—evolved into a survivor-centered evidence management tool. This shift was not due to technical barriers, but ethical insight gained through interdisciplinary reflection and survivor engagement.

RQ prompted us to question core assumptions: Why prioritize detection? Who is helped—or harmed—by such tools? These inquiries revealed the risks of over-surveillance, false positives, and retraumatization. Engaging with survivors reoriented our goals from system enforcement to individual autonomy. We moved from detection to documentation; from surveillance to support. This pivot demanded humility and accountability. We established a survivor-led advisory board and adopted trauma-informed engagement practices. RQ revealed that building the “right” AI tool sometimes means not building the original one at all.

Though born from HT, RQ is domain-agnostic. Its five steps can guide AI projects in other high-stakes areas like child welfare or predictive policing, where social complexity and contested ethics are common. However, its value lies not in universal questions but in enabling domain-specific deliberation. RQ is not a checklist—it’s a scaffold for relational ethics, adaptable to differing legal, cultural, and moral contexts.

RQ repositions ethics as foundational—not auxiliary—to design. It challenges the dominant logic of “build fast, comply later,” advocating instead for a stance of relational responsibility. In domains where harm is obscured by urgency and moral righteousness, such reflection is not optional—it’s essential.

From our experience, we offer the following recommendations for researchers and practitioners considering the RQ approach:
1) Reframe ethical inquiry as an entry point, not an afterthought. Ethical reflection should begin before any AI design work—not after feasibility is established. 2) Institutionalize the option of not building. Ethical frameworks should allow space to walk away from a project if harms outweigh benefits. 3) Replace fixed checklists with iterative dialogue. Tools like RQ must remain flexible and open-ended, prioritizing critical reflection over formal compliance. 4) Prioritize relational accountability. Establish advisory structures or partnerships with affected communities that extend throughout the project lifecycle and not just during initial consultation. 5) Rethink success metrics. In sensitive domains, success may mean empowerment, harm reduction, or withdrawal—not optimization or scale. 6) Ultimately, RQ fosters a philosophy of deliberate ethics: one in which developers embrace uncertainty, remain open to critique, and recognize that responsible AI is not merely a matter of doing things right—but asking if they should be done at all.

\bibliographystyle{ACM-Reference-Format}
\bibliography{acmart}

\end{document}